\documentclass{article}

\usepackage{arxiv}

\usepackage[utf8]{inputenc} % allow utf-8 input
\usepackage[T1]{fontenc}    % use 8-bit T1 fonts
\usepackage{hyperref}       % hyperlinks
\usepackage{url}            % simple URL typesetting
\usepackage{booktabs}       % professional-quality tables
\usepackage{amsfonts}       % blackboard math symbols
\usepackage{nicefrac}       % compact symbols for 1/2, etc.
\usepackage{microtype}      % microtypography
\usepackage{lipsum}
\usepackage{graphicx}

\title{The Second Worldwide Wave of Interest in Coronavirus since the COVID-19 Outbreaks in South Korea, Italy and Iran: A Google Trends Study}

\author{
  Artur Strzelecki \\
  Department of Informatics\\
  University of Economics in Katowice\\
  40-287 Katowice, Poland \\
  \texttt{artur.strzelecki@ue.katowice.pl} \\
}

\begin{document}
\maketitle

\begin{abstract}
\textit{Objective}: The recent emergence of a new coronavirus, COVID-19, has gained extensive coverage in public media and global news. As of 24 March 2020, the virus has caused viral pneumonia in tens of thousands of people in Wuhan, China, and ten of thousands of cases in 184 other countries and territories. This study explores the potential use of Google Trends (GT) to monitor worldwide interest in this COVID-19 epidemic. GT was chosen as a source of reverse engineering data, given the interest in the topic.

\textit{Methods}: Current data on COVID-19 is retrieved from (GT) using one main search topic: Coronavirus. Geographical settings for GT are worldwide, China, South Korea, Italy and Iran. The reported period is 15 January 2020 to 24 March 2020.

\textit{Results}: The results show that the highest worldwide peak in the first wave of demand for information was on 31 January 2020. After the first peak, the number of new cases reported daily rose for 6 days. A second wave started on 21 February 2020 after the outbreaks were reported in Italy, with the highest peak on 16 March 2020. The second wave is six times as big as the first wave. The number of new cases reported daily is rising day by day.

\textit{Conclusion}: This short communication gives a brief introduction to how the demand for information on coronavirus epidemic is reported through GT.

\end{abstract}

% keywords can be removed

\section{Introduction}
As of 7 pm Central European Time on 24 March, 2020, 407,485 cases of pneumonia had been reported globally \cite{Chen2020,Cheng2020,Hui2020} that were caused by the novel coronavirus that is now known as COVID-19 \cite{WorldHealthOrganization2020c}. 18,227 cases have resulted in death \cite{Parry2020a}. There have been 325,894 reported cases in other 184 countries and territories, including Italy (n=69,176), Iran (n=24,811) and South Korea (n=9,037) \cite{Nishiura2020,Dong2020}. It took over three months to reach the first 100 00 confirmed cases, 12 days to reach the next 100 000, 4 days to reach the next 100 000 and only 3 days to reach the next 100 000 \cite{WorldHealthOrganization2020a}.

On 11 February 2020, the official names were announced for the virus responsible for COVID-19 (previously known as “2019 novel coronavirus” or “2019-nCoV”) and the disease it causes. The official name of the disease is coronavirus disease (COVID-19); the virus itself is called severe acute respiratory syndrome coronavirus 2 (SARS-CoV-2) \cite{WorldHealthOrganization2020c}. Thus, the aim of the study is to show how GT data can be used to forecast the trends of reporting new cases.

This article can be a starting point for further analysis of information demand across search engine. Currently, Google offers its Trends service, which acts as reverse data engineering and allows data on users’ searches to be collected, which in this case is interest in the COVID-19 epidemic.

\section{Materials and methods}
The methodology for this communication follows the principles presented by \cite{Mavragani2019} that describe how to select the appropriate keyword(s), region(s), period, and category. Data is collected from GT and is normalized. High interest in a search query is expressed by 100, whereas a lack of interest or insufficient data is expressed by 0. GT contains data from different geographical locations that is segmented into countries, territories and cities; it also allows a custom time range to be set.

Data for the study was retrieved for the period starting 15 January 2020, as this was when relevant data started appearing on GT. The data comes from a textual search with five geographical settings: 1) worldwide, to see the global interest in coronaviruses; 2) China, where there is currently the highest number of cases; 3) South Korea, where interest has increased since 19 February because hundreds of new cases were reported; 4) Italy; and 5) Iran, where since 22 February hundreds of new cases have been reported. Data from GT related to interest in coronavirus was compared with confirmed reports of new cases provided by WHO \cite{WorldHealthOrganization2020a}. The collected data relates to the search topic: Coronavirus. This topic allows the popularity of all related keywords across all available languages and regions to be compared \cite{Kaminski2019}.

\section{Results}
On Google Trends, the first wave of interest in coronavirus peaked on 31 January 2020. This is measured globally for all GT data for the coronavirus search topic. Since 1 February, global interest has decreased even though the number of new cases reported daily is increasing. In the first wave, the highest number of confirmed new cases was on 5 February. In the second wave, so far the highest number of confirmed new cases in a single day was on 24 March.

\begin{figure}[hbt!]
    \centering
    \includegraphics[width=\textwidth]{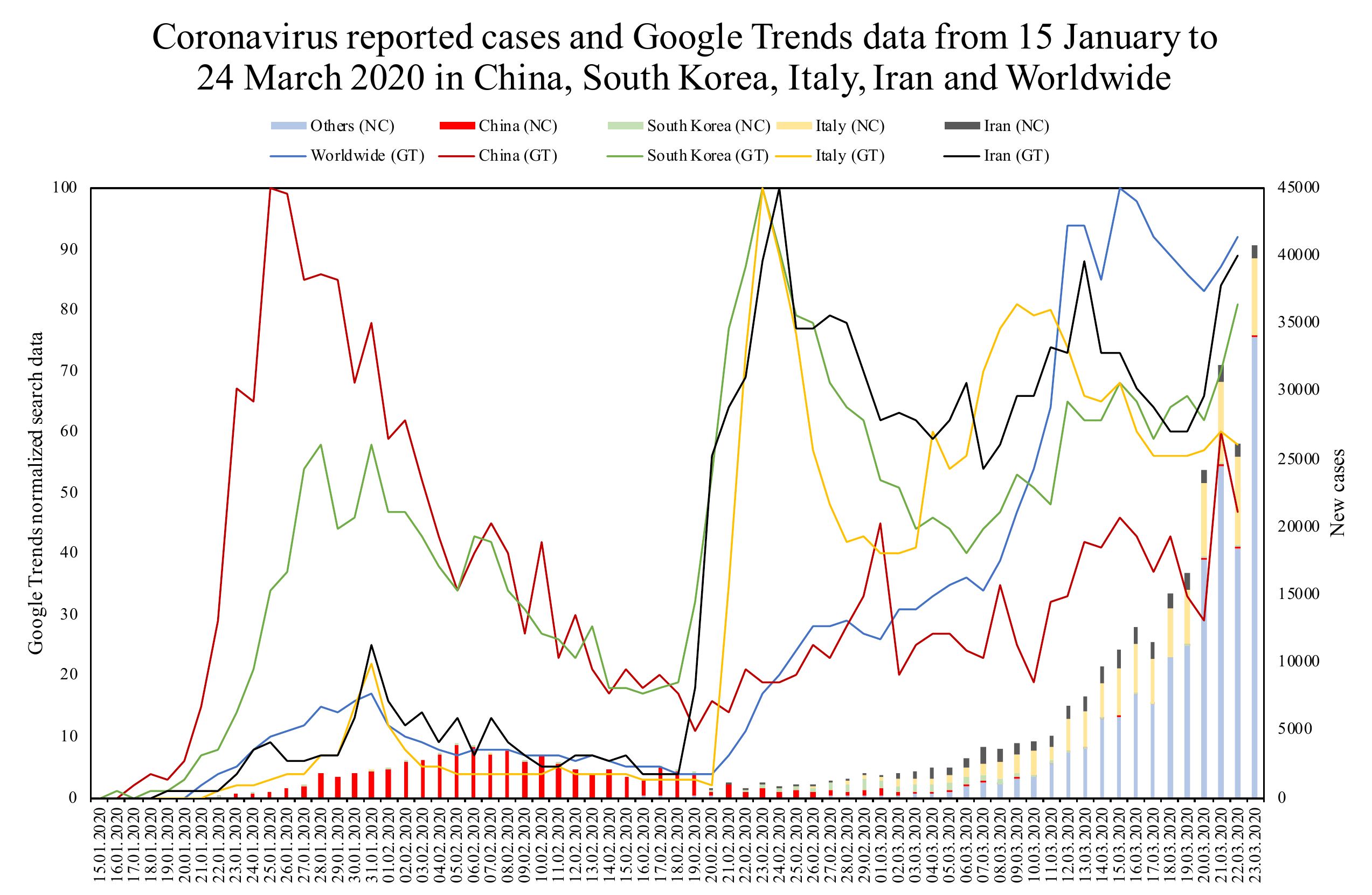}
    \caption{Coronavirus reported cases and Google Trends data from 15 January to 24 March 2020 in China, South Korea, Italy, Iran and Worldwide.}
    \label{fig:fig1}
\end{figure}

Figure \ref{fig:fig1} presents the global, Chinese, South Korean, Italian and Iranian results compared to the number of new COVID-19 cases. Since GT has a two-day data delay, GT data at the time of writing ends on 21 March. The left axis shows normalized GT search volume. The right axis shows new COVID-19 cases \cite{WorldHealthOrganization2020c}. The data interval is one day.

The situation has changed since a rapid increase in cases was reported in South Korea, Italy and Iran. GT data reveals the rapid growth of the second wave of interest in coronavirus since 21 February 2020. This rising interest trend is observed worldwide and in the presented countries, where a rapid increase in cases of laboratory-confirmed COVID-19 has been reported since 21 February 2020 \cite{WorldHealthOrganization2020b}.

\begin{table}
 \caption{Pearson correlation matrix of GT trends between China, South Korea, Italy, Iran and Worldwide from 15 January to 18 February 2020 - the first wave}
  \centering
\begin{tabular}{llllll}
    \toprule
 & China & South Korea & Italy    & Iran     & Worldwide\\
    \midrule
China       & 1           &          &          &           &   \\
South Korea & 0.762***    & 1        &          &           &   \\
Italy       & 0.348***    & 0.647*** & 1        &           &   \\
Iran        & 0.575***    & 0.780*** & 0.817*** & 1         &   \\
Worldwide   & 0.779***    & 0.890*** & 0.781*** & 0.670     & 1 \\
    \bottomrule
\multicolumn{6}{l}{* \textit{p}<0.1, **\textit{p}<0.05, ***\textit{p}<0.01}
\end{tabular}
  \label{tab:table1}
 
\end{table}

\begin{table}
 \caption{Pearson correlation matrix of GT trends between China, South Korea, Italy, Iran and Worldwide from 19 February to 24 March 2020 - the second wave}
  \centering
\begin{tabular}{llllll}
    \toprule
 & China & South Korea & Italy    & Iran     & Worldwide\\
    \midrule
China       & 1           &          &          &           &   \\
South Korea & 0.072***    & 1        &          &           &   \\
Italy       & 0.171***    & 0.433 & 1        &           &   \\
Iran        & 0.309***    & 0.714*** & 0.596*** & 1         &   \\
Worldwide   & 0.777***    & 0.047** & 0.297* & 0.246***     & 1 \\
    \bottomrule
\multicolumn{6}{l}{* \textit{p}<0.1, **\textit{p}<0.05, ***\textit{p}<0.01}
\end{tabular}
  \label{tab:table2}
 
\end{table}

Table \ref{tab:table1} presents Pearson’s correlation matrix of GT trends between Worldwide and in reported countries in the time of first wave, from 15 January to 18 February 2020. Results show that Worldwide trend is significantly, positively and strongly correlated with trends in China, South Korea and Italy.

Table \ref{tab:table2} presents Pearson’s correlation matrix of GT trends between Worldwide and in reported countries in the time of second wave, from 19 February to 24 March 2020. Results show that Worldwide trend is significantly, positively correlated with trends in China, Italy, South Korea and Iran.

\section{Discussion}
The key finding is that GT forecasted the rise of new cases. In first wave, new cases increased day-by-day for 6 days after the highest peak of GT worldwide interest. In the second wave, interest in coronavirus on GT is still rising, which predicts the increasing number of new cases reported daily. This implies that national health services should implement additional health measures against countries other than China.

The limitation of the study is GT data about China because of the general unavailability of Google in China. Another limitation is that data about South Korea, Italy and Iran is changing rapidly every day, thus results are only relevant to the reported date.

%\bibliographystyle{IEEEtran}  
%\bibliography{bibtex}  %%% Remove comment to use the external .bib file (using bibtex).

%

\end{document}